\documentclass[12pt, a4paper]{article}
\usepackage{amsmath}
\usepackage{color, xcolor}
\usepackage{graphicx}
\usepackage{pdfpages}
\usepackage{geometry}
\usepackage{caption}
\usepackage{lineno}
\usepackage{authblk}
\usepackage{float}
\usepackage{cite}

\title{Multimodes of Rayleigh wave excited by hurricane Dorian}

\author{Xuping Feng \thanks{geophydogvon@gmail.com} }
\affil{Department of Earth and Space Sciences, Southern University of Science and Technology, Shenzhen, Guangdong, 518000, China}

\date{}

\begin{document}

\maketitle

\section*{keypoints}
	$\bullet$ Hurricane Dorian has continuously excited signals recorded by seismic stations in the southeast of the United States from 4th to 6th Sept 2019
	
	$\bullet$ EGFs with high SNRs were extracted by cross-correlating ambient seismic noise released by Dorian for several days or even a single day
	
	$\bullet$ Clear multimodal dispersion curves (3$\--$4 modes) of Rayleigh waves emerged from retrieved EGFs
\clearpage


\section*{Abstract}
	Extracting Empirical Green's Functions (EGFs) by cross-correlating ambient seismic noise or coda, considered as an efficient approach of retrieving new response inside the media of two receivers, has been developed well in the past two decades. However, EGFs emerge with high Signal-to-Noise Ratios (SNRs), which needs us to stack a large number of cross-correlations. Here, we retrieve EGFs with high SNRs from ambient seismic noise released by hurricane Dorian moving along the eastern coast of the United States during 3 days in Sept 2019. We systematically analyze the energy sources of seismic noise records with time-frequency analysis and beam-forming. We find that the energy exhibits very strongly when hurricane Dorian progresses along the coast from 4th to 6th Sept. Our results suggest that EGFs exhibit strong directionality and only emerge in the negative part of cross-correlation time functions due to the noise sources of Dorian. We extract clear multimodal dispersion curves of Rayleigh waves from retrieved EGFs. Our results, indeed, reveal that multimodal dispersion curves pf Rayleigh wave could be extracted from ambient seismic noise excited by Dorian for several days or even a single day, which means that we enlarge the seismic data sets for the furture work of investigating shear velocity structures of the Earth's crust and upper mantle and understanding our planet more deeply.

\section*{Plain Language Summary}
	We detect the surface wave coherent signals cross-correlating ambient seismic noise excited by hurricane Dorian since there is no clear signal generated by hurricanes like earthquakes. Therefore, cross-correlating noise enlarges the useful data sets, especially these poorly sampled areas where there are no larger earthquakes occuring. So this helps us illuminate the Earth's interiors much better and further understand the evolution of our planet.

\section{Introduction}
	Benifited from the pioneering ideas of \cite{Aki1957} and \cite{Claerbout1968}, interstation cross-correlations from ambient seismic noise or coda is an efficient approach of retrieving new response inside the media under the station pair \cite{Shapiro2005high, Yang2007ambient, Yao2006Surface, Campillo2003long, Wang2015equatorial}. Ambient seismic noise or coda cross-correlation has far-reaching implications for the field of passive seismic imaging and surface wave speeds up the applications to illuminating the Earth's velocity structures from reginal to continental scales \cite{Bakulin2006virtual, Campillo2003long, Yang2007ambient, Yao2006Surface, Moschetti2007surface, Lin2008surface}.

	However, extracting dispersion curves from EGFs by ambient noise needs large data sets of long duration since we do not fully understand how the noise sources distribute. Events, such as earthquakes, must have high SNRs in order to retrieve clear dispersion curves. Furthermore, multimodal dispersion curves have stronger constraints on the shear wave velocity model than the fundamental mode \cite{Xia2003inversion, Pan2019sensitivity} .

	Here, we cross-correlate vertical-component ambient seismic noise continuously excited by hurricane Dorian along the southeastern coast of the United States for 3 days in Sept 2019 and extract Rayleigh surface wave EGFs with high SNRs. We analyze the energy spectra by time-frequency analysis and constrain the distributions of noise sources using beam-forming array method \cite{Rost2002array}. Our results suggest that Dorian locates in the direction of the alignment of the station pairs, which leads us to retrieve clear multimodal dispersion curves of Rayleigh wave from stable EGFs.

\section{Data and Methodology}
	In this study, we do not apply any other different techniques but follow the ambient seismic data processing procedures suggested by \cite{Bensen2007processing}. We only calculate cross-correlations of stations in array CO with those in array ET. For more details, (1) pre-processing and cutting data into segments and every segment is with a length of 4 hours; (2) doing band-pass filtering of 0.05$\--$0.495 Hz and executing temporal normalization with running absolute mean method; (3) spectral whitening with the same algorithm in step (2); (4) computing cross-correlation functions and temporal stacking all segments; (5) selecting EGFs with high SNRs by setting a threshold.
	
	We calculate the SNRs of Rayleigh waves according to the following steps: (1) giving a rough reference phase velocity of Rayleigh waves $V_{ref}$, we calculate the theoretical travel time $t_{ref}$ with $V_{ref}$ and interstation dictance $d$ using $t_{ref}=\frac{d}{V_{ref}}$; (2) setting a half window $t_{h}$ with a length of 15 seconds to search the maximal absolute amplitude of Rayleigh wave, $P_{ray}$, and $P_{ray} =|A_k|_{max}$; (3) choosing a noise window with lag time from -1000 to -500 seconds since there is no significant signal in this window and calculating the standard deviation $V_{std}$ of the absolute amplitute in this window (Figure S1) ; (4) lastly, we define the SNR of Rayleigh waves $SNR_{ray}$ as follows
	
	\begin{equation}
	SNR_{ray} = \frac{P_{ray}}{V_{std}}.
	\end{equation}
	
	For the extraction of Rayleigh surface waves, we use MASW (Multichannel Analysis of Surface Wave) technique, phase shift \cite{Park1998imaging} to obtain the dispersion images and this method follows the forluma,
	\begin{equation}
	P(w, k) = \int_{-\infty} ^{\infty}\frac{U(\omega, x_n)}{|U(\omega, x_n)|}e^{ikx_n}dx,
	\end{equation}
	where, $\omega$ is angular frequency, $k$ is wavenumber, and $U(\omega, x_n)$ is the Fourier spectrum of displacement $u(t, x_n)$ recorded by the $n_{th}$ staion at position $x_n$. For the directness of dispersion images of Rayleigh waves, using the relation $k=\frac{2\pi f}{c}$, we rewrite eq. $(2)$ in discrete form as
	\begin{equation}
	P(f_l, c_j)=\sum_{n=1}^{N}\frac{U(f_l, x_n)}{|U(f_l, x_n)|}e^{\frac{i2\pi f_l}{c_j}x_n}\Delta x,
	\end{equation}
	where, $N$ is the total number of seismic stations.
	
	Continuous seismic vertical-component data used in this study are from 4th to 6th Sept 2019 and recorded by stations in arrays CO and ET installed in the east of the United States (Figure 1). However, we processed more data of station CSB in array CO and MGNC in array ET to execute time-frequency analysis \cite{Sejdic2009time} for comparing the energy distribution before and during the period of hurricane Dorian moving along the coast.

\section{Results and Discussion}
	\subsection{The Energy Contribution to EGFs Construction}
	We clip the displacement waveform under an absolute amplitude threshold equal to $1.5e^{-5}$ meters to clearly show the hurricane event. As we can see that there is a strong envelope instead of a clear first arrival (Figure 2a, Figure S2), and that is the reason why we consider these signals as ambient noise to retrieve EGFs. The output spectra of time-frequency analysis applied to displacement waveform recorded by station CSB in array CO suggest that Dorian arises and vanishes on Sept 1st and 8th, respectively. The spectra especially emerge more strongly from the waveform during Sept 4th through 6th and exhibit the most powerfully on Sept 5th (Figure 2b, Figure S2). This phenomenon is well in agreement with the EGFs and dispersion curves of Rayleigh waves extracted from this duration discussed in the next part.
	
	Furthermore, our results reveal that the EGFs only significantly emerge in the negative part of the cross-correlation time functions. Indeed, the signals in negative and positive parts are asymmetrical, or there exist phase shifts in cross-correlations when the noise sources do not evenly distribute \cite{Gouedard2008convergence, Yao2009analysis}. Here, the cross-correlation function in this study $R_{xx}(t)$ is given by
	\begin{equation}
	R_{xx}(t) = \int_{-\infty}^{\infty}u_{ET}(\tau)u_{CO}(\tau+t)d\tau,
	\end{equation}
	where, $u_{CO}(t)$ and $u_{ET}(t)$ are the displacements recorded by stations in arrays CO and ET, respectively. However, the Rayleigh waves emerge in the negative part but no significant signal in the positive part of cross-correlation (Figure 2d), which indicates that the noise sources almost lie on the same great circle arc determined by the station pair to mainly contribute to reconstructing EGFs (Figure 2c). And this phenomenon is theoretically and practically reported in some researches \cite{Snieder2004extracting, Gouedard2008convergence, Snieder2015Seismic, Feng2019}. More specifically, the noise source needs to locate on the extension line from array ET to CO. Here we use beam-forming array method \cite{Capon1970applications, Rost2002array} to constrain the energy distributions from 4th to 6th Sept. We execute beam-forming analysis with data recorded by array ET in these 3 days and the frequency band and slowness range are set as 0.05$\--$0.495 Hz and 20$\--$40 seconds/degree (velocity 2.78$\--$5.56 km/sec), respectively. Eventually, the directions and velocity range of noise sources match well with the evolution path of Dorian (Figure 3). Furthermore, the positions of Dorian in these 3 days agree well with the alignment of station pairs in arrays CO and ET during this period. Therefore, the results reveal that hurricane Dorian mainly contributes to the emergence of clear Rayleigh waves (here vertical components used) extracted from ambient seismic noise.
	
	\subsection{Multimodal Dispersion Curves of Different Days}
	The extraction of dispersion curves is the key to imaging the shear wave velocity structures and understanding the evolution of the interiors of the Earth \cite{Li2020effective}. Multimodal dispersion curves can constrain the shear wave velocity model much better than only using fundamental mode \cite{Xia2003inversion, Pan2019sensitivity, Wu2020shear}. We constrain the directions of noise sources, Dorian, using beam-forming, and the sources restrictedly locate in the directions close to alignment of station pairs, which means that Dorian is the stationary phase point of reconstructing the EGFs from ambient noise \cite{Snieder2004extracting, Tonegawa2009seismic}. This reveals that the Raleigh wave EGFs retrieved emerge stably and we can extract dispersion curves from these EGFs. Here we use retrieved EGFs by stacking all 6 segments with a length of 4 hours for every day from 4th to 6th Sept (Figure 4a$\--$c). We set an SNR criterion of 6.0 for obtaining high-quality EGFs, and the numbers of EGFs with SNRs greater than or equal to 6.0 are little different since Dorian deviates from the arc determined by station pairs on 4th and 6th Sept. Furthermore, the magnitude of Dorian is indeed larger on 5th Sept than those on 4th and 6th Sept. Therefore, the dispersion curves of Rayleigh waves have 4 modes, including the fundamental and 3 higher overtones on 5th Sept. However, there are only 3 modes on 4th and 6th Sept (Figure 4d$\--$f, Figure S3d-f).

	\subsection{Continuous excitations for retrieving EGFs}
	Stacking a large number of temporal cross-correlations is the key to constructing stable EGFs from ambient seismic noise, and in general, acquisition of 1 to 2 years of ambient noise is sufficient \cite{Bensen2007processing}. However, we do not need so many noise sources or data of long duration if we do know the distributions of them locating on the same arc with station pairs. \cite{Takagi2006rayleigh} imaged the velocity structures with group velocity of Rayleigh wave under the whole region of Japan cross-correlating noise released by 4 typhoons. Here, Dorian lying on the alignment of station pairs has continuously been exciting the energies for emergence of stable EGFs, which means that there are so many energy sources exciting signals for retrieving EGFs. That is the reason why we extract EGFs with such high SNRs using data recorded during a single day, and here let's say 5th Sept. \cite{Gerstoft2007year} used beam-forming and cross-correlations to analyze Rayleigh surface wave and distant body wave \cite{Gerstoft2008global} with microseism data for about 1 year, but for the seasonal variations of cross-correlations, we need to large data sets in order to obtain good estimations of EGFs \cite{Yao2009estimation}. Fortunately, there are some events except earthquakes, like hurricanes, that we haven't utilized well to extract EGFs, which offer us more opportunities to illuminating the interiors of our Earth.
	
\section{Concluding Remarks}
	Good estimations of EGFs extracted from ambient seismic noise excited by hurricane Dorian suggests that clear multimodal dispersion curves of Rayleigh waves, here, 3$\--$4 modes, can emerge. Unlike noise excited by unknown earthquakes, we use data of several days or even a single day to extract EGFs with high SNRs. There are plenty of storms, especially strong hurricanes that excite the energies propagating through the media and provide rich information about the interiors of our Earth. More importantly, the data of retrieving stable EGFs from storms are less than those from earthquakes since we know how the storms distribute. Indeed, making the most of the signals generated by storms recorded by seismic stations may lead us to explore the shear wave velocity model of the Earth's crust and upper mantle much better and understand the evolution of our planet more deeply.

\section{acknowledgments}
	The IRIS services, especially the IRIS Data Management Center gave us the access to seismic recordings used in this study. This work was supported by Natural Science Foundation of China.

\bibliographystyle{plain}
\bibliography{Ref}

\begin{figure}[H]
	\centering  
	\includegraphics[scale=0.4]{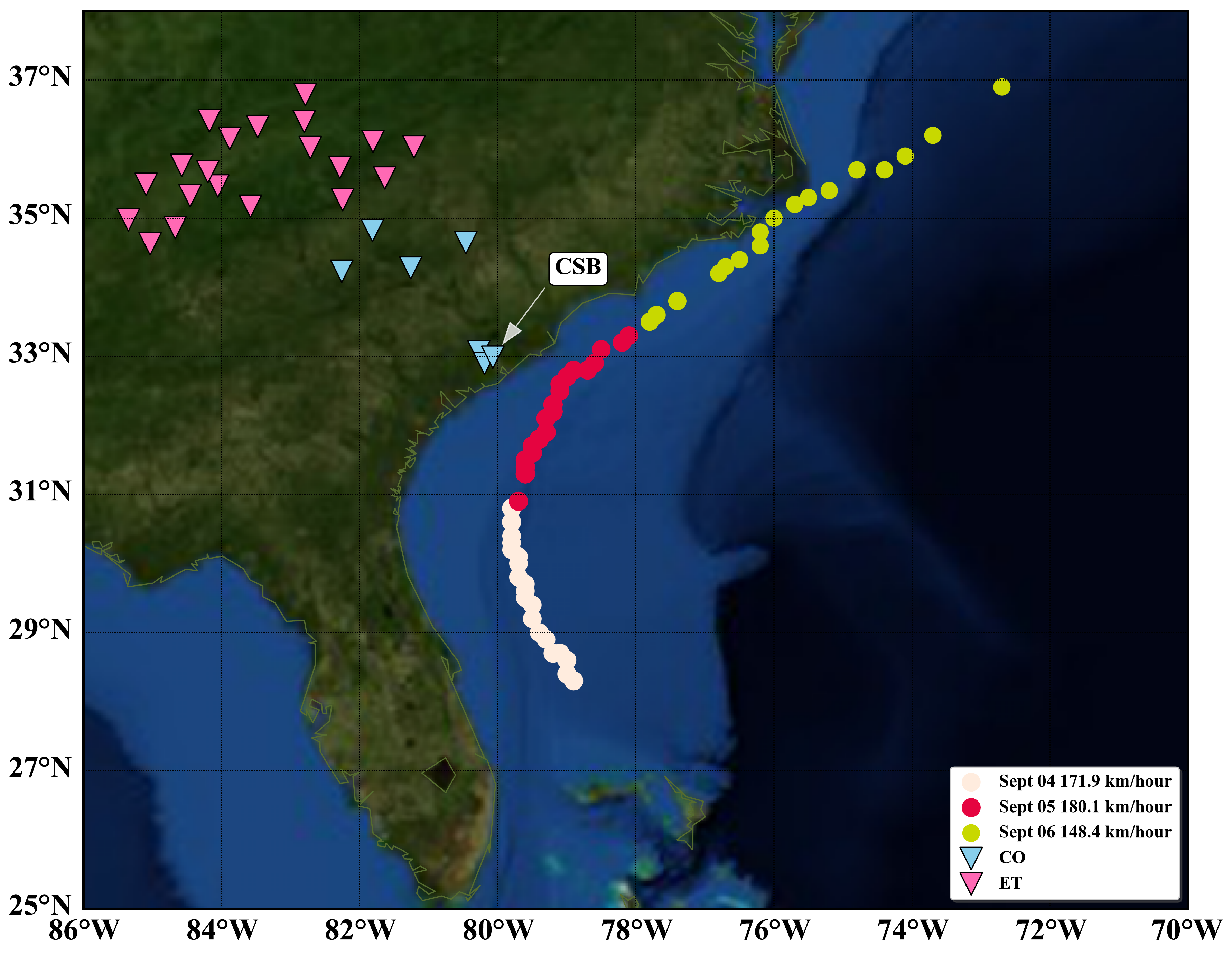}  
	\caption{Seismic stations in arrays CO (light blue solid triangles) and ET (light red solid triangles) in the southeast of the United States, the position evolution of Dorian (solid circles) during Sept 4th through 6th 2019 and the sizes of solid circles denote the traveling speed of Dorian. Waveform recorded by station CSB indicated by white arrow will be used to do time-frequency analysis.}  
	\label{1}  
\end{figure}

\begin{figure}[H] 
	\centering  
	\includegraphics[scale=0.42]{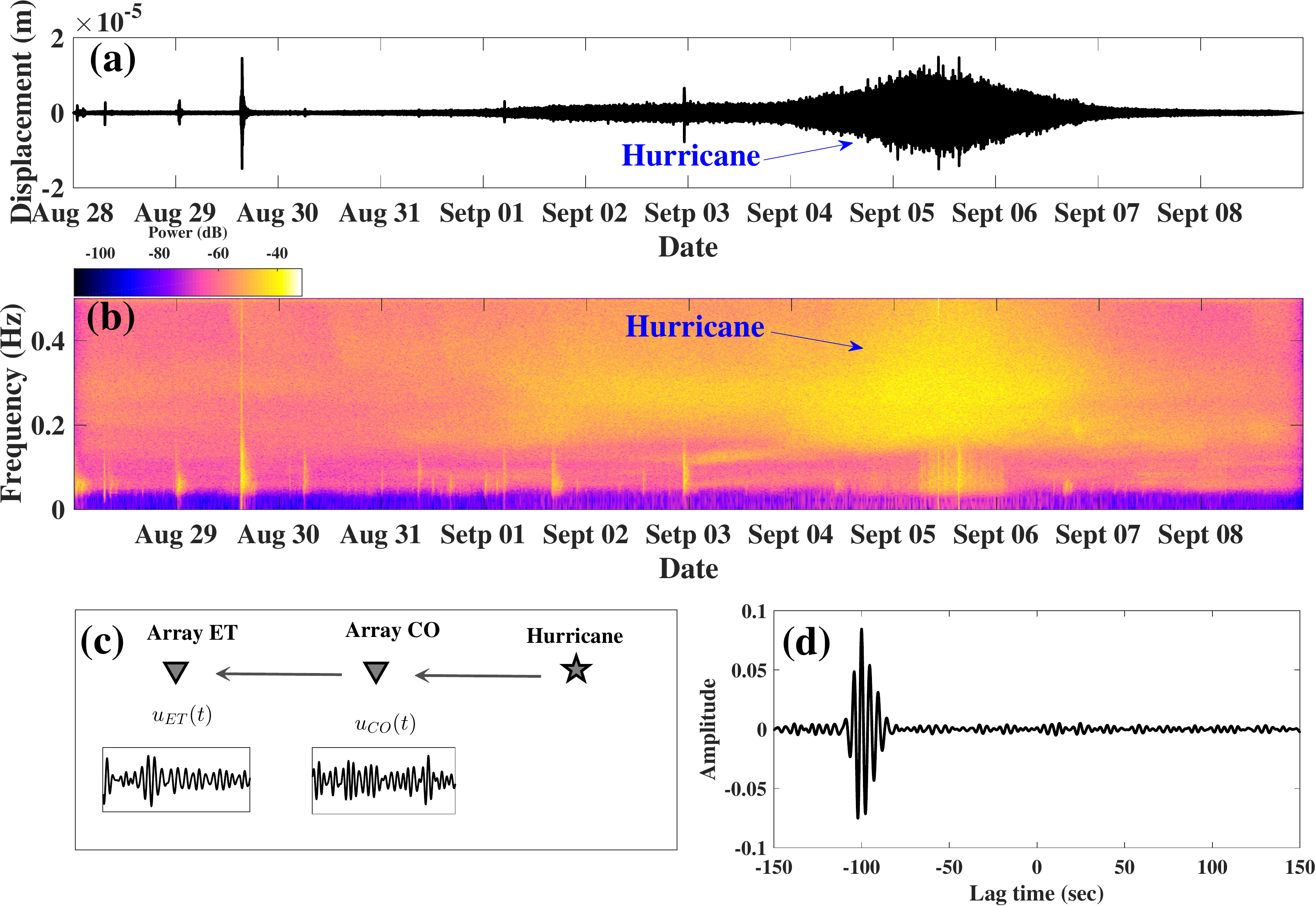}  
	\caption{The displacement waveform, corresponding time-frequency spectra and cross-correlations sketch. (a) The displacement waveform recorded by staion CSB in array CO. (b) Time-frequency spectra of displacement waveform. The blue arrows indicate the strongest energy excited by Dorian recorded by CSB during Sept 4th to 6th 2019. (c) The propagation direction of energy generated by Dorian. (d) One cross-correlation time function computed by two stations separately in arrays CO and ET on Sept 5th 2019 and Rayleigh wave only emerges in the negative part.}  
	\label{2}  
\end{figure}

\begin{figure}[H] 
	\centering  
	\includegraphics[scale=0.475]{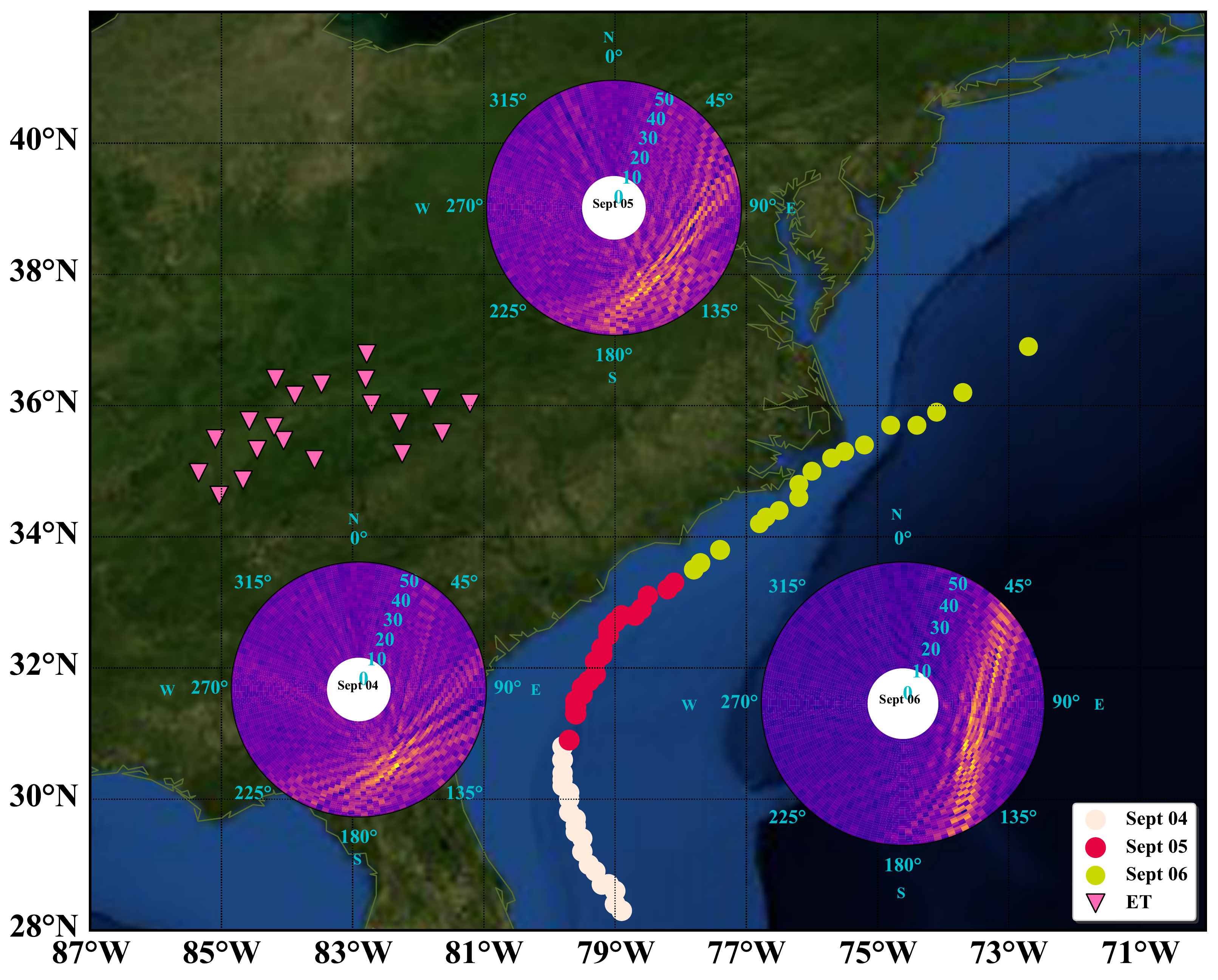}  
	\caption{The outputs of beam-forming analysis with displacement waveform data recorded by stations in array ET (light red solid triangles) during 4th through 6th Sept 2019 (the texts on centers of polar plots denote the date number of these 3 days and yellow bins on polar plots indicate the significant energies of Rayleigh waves) and the tracking path of Dorian (solid circles).}  
	\label{3}  
\end{figure}

\begin{figure}[H] 
	\centering  
	\includegraphics[scale=0.38]{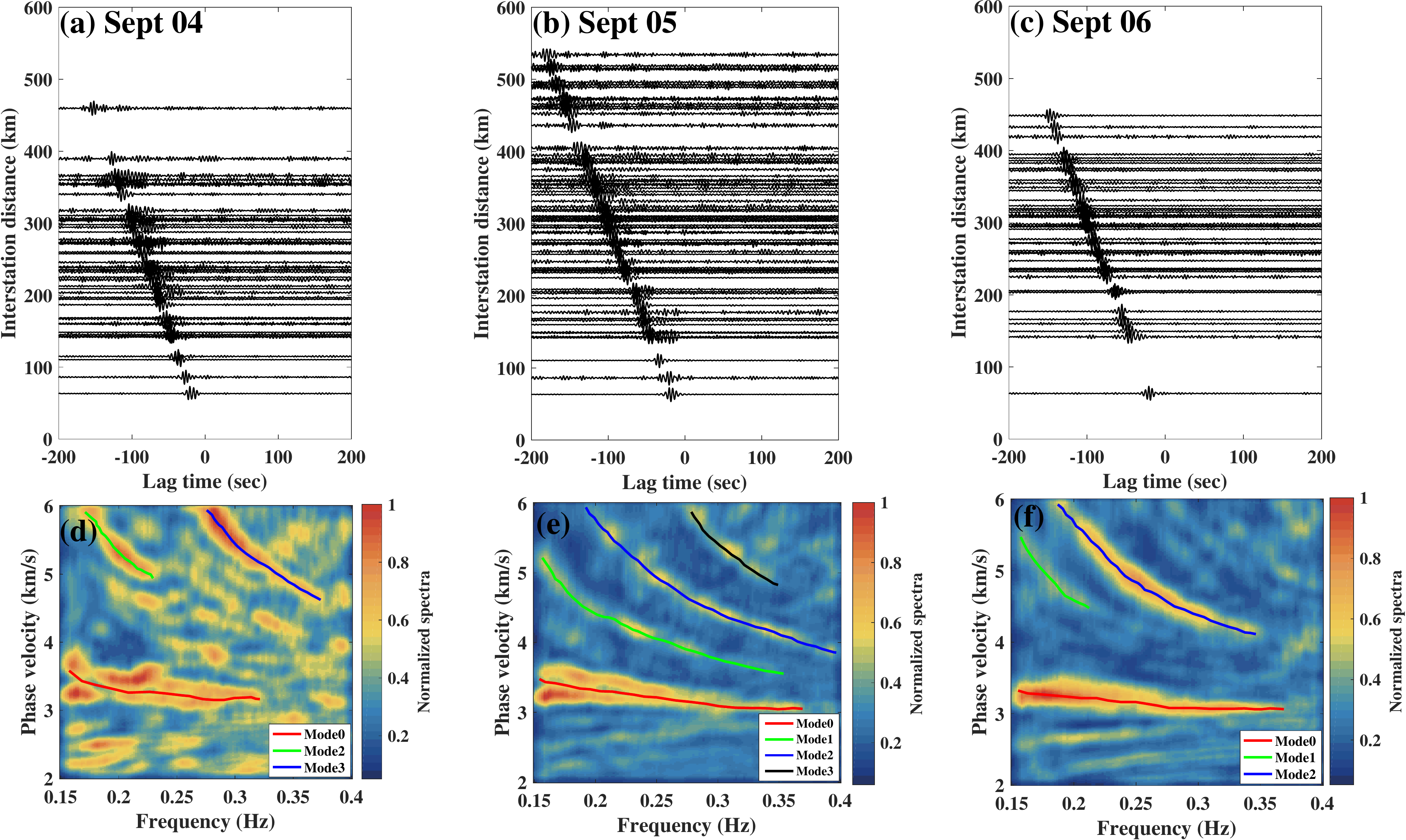}  
	\caption{Retrieved EGFs and dispersion curves of Rayleigh waves extracted from corresponding EGFs. (a$\--$c) EGFs retrieved from ambient noise recorded by stations in arrays CO and ET on 4th to 6th Sept 2019, respectively. (d$\--$f) Dispersion images of Rayleigh waves separately extracted from EGFs on panels (a$\--$c) and manually picked dispersion curves (denoted by solid lines).}
	\label{4}  
\end{figure}

\end{document}